\documentclass[10pt,aps,pra,twocolumn,longbibliography]{revtex4-1}
\usepackage{amsmath}
\usepackage{amssymb}

\newcommand\beq{\begin{equation}}
\newcommand\eeq{\end{equation}}
\newcommand\bea{\begin{eqnarray}}
\newcommand\eea{\end{eqnarray}}
\newcommand\nn{\nonumber}

{}

\newcommand\tr{\mathsf{tr}}

\newcommand\ox{\otimes}

\newcommand\ro{\hat{\rho}}
\newcommand\sio{\hat{\sigma}}

\newcommand\al{\alpha}

\begin{document}
\title{Operational meaning of the classical fidelity and the path length in Fisher-Kubo-Mori-Bogoliubov geometry} 
\author{Lajos Di\'osi}
\email{diosi.lajos@hun-ren.wigner.hu}
\homepage{www.wigner.hu/~diosi} 
\affiliation{Wigner Research Center for Physics, H-1525 Budapest 114 , P.O.Box 49, Hungary}
\affiliation{E\"otv\"os Lor\'and University, H-1117 Budapest, P\'azm\'any P\'eter stny. 1/A}
\date{\today}

\begin{abstract}
We show that the minimum entropy production in near-reversible quantum state transport
along a path is simple function of the  path length measured according to the Fisher--KMB metrics. 
Hence the sharp values of path lengths, also called statistical lengths, 
obtain operational meaning to quantify the residual irreversibility in near-reversible
state transport. 
In the classical limit, the Bhattacharryya fidelity obtains a sharp operational meaning
after eighty years.
\end{abstract}
\maketitle

\section{Introduction}
A quantitative comparison between two probability distributions $p_\al$ and $q_\al$ 
can be based on the Bhattacharyya coefficient published in 1945 \cite{bhattacharyya1943measure}: 
\beq\label{FC}
F(p,q)=\sum_\al \sqrt{p_\al q_\al}~\in[0,1].
\eeq 
This function defines a measure $F\in[0,1]$ of closeness between $p$ and $q$.  
Fifty years later, to measure the closeness of two quantum states $\ro$ and $\sio$,
Jozsa borrowed the expression known in mathematics 
\cite{bures1969extension,uhlmann1976transition} and called
it fidelity \cite{jozsa1994fidelity}:
\beq\label{FQ}
F(\ro,\sio)=\Vert\sqrt{\ro}\sqrt{\sio}\Vert_{\tr}=\tr \sqrt{\sqrt{\sio}\ro\sqrt{\sio}}~\in[0,1].
\eeq
Quantum fidelity is the central tool in modern quantum informatics \cite{nielsen2001quantum}.
In the special (classical) case when $[\ro,\sio]=0$, we can write  
$\ro=\mathsf{diag}[p_1,p_2,\dots,p_\al,\dots]$ and  
$\sio=\mathsf{diag}[q_1,q_2,\dots,q_\al,\dots]$. 
The quantum fidelity \eqref{FQ} coincides with the classical fidelity \eqref{FC}.

One of our goals is the operational interpretation of $F(q,p)$ because it is not known \cite{nielsen2001quantum}. For the quantum fidelity $F(\ro,\sio)$
a simple operational interpretation was already proposed \cite{dodd2002simple}, 
based on Uhlmann's theorem \cite{uhlmann1976transition}, and goes like this.
\emph{Assume that $\ro$ and $\sio$ are reduced
states of two pure states, respectively, in a larger system. Assume that we have 
access to this larger system. Then the distinguishability of $\ro$ and $\sio$ is at least as
high as the distinguishability of two arbitrary pure states whose overlap (modulus of scalar product)
is given by the fidelity $F(\ro,\sio)$. } 
This interpretation assumes the operation on the larger system where 
both $\ro$ and $\sio$ are pure. Various theorems contain the sharp value of fidelity
but their meaning should always include operations on the `environmental' quantum system.

Fidelities are related to distances, also called 
statistical distances, 
on the underlying Riemannian geometries. 
The infinitesimal distance $d\ell$ defines 
the Fisher metrics \cite{fisher1925theory} between the probability distributions $p$ and $p+dp$
and the Bures metrics  \cite{bures1969extension} between the quantum states 
$d\ro$ and $\ro+d\ro$:   
\bea
\label{dlF}
(d\ell)_{Fisher}^2&=&2\sum_\al(d\sqrt{p_\al})^2=\sum_\al\frac{(dp_\al)^2}{p_\al},\\
\label{dlB}
(d\ell)_{Bures}^2&=&2\tr (d\sqrt{\ro})^2=\tr \left(d\ro\frac{2}{\ro_L+\ro_R}d\ro\right).                                           
\eea
The labels $L/R$ mean $\ro_L$ acts from the left, as usual, while $\ro_R$ acts from the right. 
The local Fisher metrics is the special case of the local  Bures metrics when $[\ro,d\ro]=0$.
Consider the geodesics between $p$ and $q$ or between $\ro$ and $\sio$. 
The geodesic distances $\ell$ are functions of the respective
 fidelities:
\bea
\label{Fl}
\ell(p,q)_{Fisher}&=&2\arccos F(p,q)~\in[0,\pi],\\
\label{Bl}
\ell(\ro,\sio)_{Bures}&=&2\sqrt{1-F^2(\ro,\sio)}~\in[0,2].
\eea
As we said, when $\ro$ and $\sio$ commute, then $F(q,p)$ is the special case of $F(\ro,\sio)$
and the local Fisher geometry \eqref{dlF}  is the special
case of the local Bures geometry \eqref{dlB}. 
This is no longer the case with the above global distances since $\ell_{Bures}=2\sin(\tfrac12\ell_{Fisher})$. 
The Bures geodesic distance is  shorter than the Fisher geodesic distance:
The latter is the minimum length of paths but through the commuting states.   

The infinitesimal distances, i.e.: the local metrics,  have exact meaning in metrology via the famous Cram\'er-Rao
\cite{cramer1946mathematical,radhakrishna1945information} 
and quantum Cram\'er-Rao theorems \cite{braunstein1994statistical}.
\emph{Suppose an unknown state  ---or probability distributions $p$ in the classical case--- 
at the unknown length $\ell$ measured from one end of a known path. 
We perform  $N$ independent measurements on the unknown state to estimate $\ell$,
 i.e., to estimate the unknown state. If $N\rightarrow\infty$, 
 optimum measurements ensure that the
 mean squared error of estimation goes to zero as $(\Delta\ell)^2=1/N$.}    
This interpretation concerns the local geometry at one point and does not tell us
anything about non-local features like generic  Fisher--Bures distances. 

While in metrology and informatics a natural quantum generalization of the Fisher metrics is the Bures one,
in irreversible processes it is the Kubo-Mori-Bogoliubov (KMB) metrics (cf., e.g., ref. \cite{hayashi2002two}):  
\beq
\label{dlKMB}
(d\ell)_{KMB}^2=\tr\left(d\ro d\!\log\ro\right)=\tr \left(d\ro\frac{\log\ro_L-\log\ro_R}{\ro_L-\ro_R}d\ro\right).
\eeq
Unlike the expression of geodesic distance in terms of fidelity in Fisher--Bures metrics, 
no such closed expression  is known in the Fisher--KMB geometry.  
Both geometries have been missing their global interpretations but
the close relationship of the KMB metrics with von Neumann entropies will help us.
The present work proposes the operational meaning of path lengths in the Fisher--KMB metrics, which 
implies the meaning of the geodesics. 
In the special case we obtain the operational meaning of the classical Bhattacharyya fidelity as well. 
The concept is that in near-reversible transport along a path $\gamma$ between two predefined 
states $\ro,\sio$  the minimum entropy production is quantified by the length $\ell_\gamma(\ro,\sio)$.

\section{State transport by equilibrating reservoirs}
In preparation, we define the transport of  the system's initial state $\ro$ 
into the final state $\sio$ in contact with a single reservoir.
We use the reservoir model  of ref. \cite{diosi2006exact} and the theorem therein
(cf. \cite{csiszar2007limit} for the rigorous proof).

The reservoir consists  initially of $n$ independent systems in state $\sio$ each and
we start from the composite system-reservoir state  $\ro\ox\sio^{\ox n}$.
The contact of the system and the reservoir assumes a reversible step, 
a swap between the system's state $\ro$ and (e.g. the 1st) one of the reservoir's states $\sio$: 
\beq
\ro\ox\sio^{\ox n}\Rightarrow\sio\ox\ro\ox\sio^{\ox(n-1)}.
\eeq
The second step, concerning the reservoir only,  is the  irreversible relaxation of the reservoir
to a homogeneous state, modeled  by the following twirl over the permutation group: 
\beq
\ro\ox\sio^{\ox(n-1)}
\Rightarrow\frac{1}{n}\sum_{k=0}^{n-1}\sio^{\ox k}\ox\ro\ox\sio^{\ox(n-k-1)}.
\eeq
According to this model,  the transport $\ro\rightarrow\sio$ of the system state
leads to the irreversible relaxation inside the reservoir. 
In the infinite reservoir limit $n\rightarrow\infty$, fortunately, the entropy
production of relaxation can be expressed as the relative entropy 
$S(\ro\Vert\sio)=\tr(\ro\ln\ro-\ro\ln\sio)$:
\bea
&&\!\!\!\!\!S\!\left(\frac{1}{n}\sum_{k=0}^{n-1}\sio^{\ox k}\ox\ro\ox\sio^{\ox(n-k-1)}\right)
-S\!\left(\ro\ox\sio^{\ox(n-1)})\right)
\nn\\
&&~~~~~~~~\xrightarrow[n\rightarrow\infty]{}S(\ro\Vert\sio).
\eea
This is the entropy production of the single-step state transport as well:
\beq
\Delta S=S(\ro\Vert\sio).
\eeq

The  entropy production can be smaller and can even go to zero if we apply sequential 
equilibration with  intermediate reservoirs along a path, such that the stepsizes go to zero.  
The following feature of relative entropy will be important \cite{petz1993bogoliubov}:
\beq\label{relSdell}
S(\ro\vert\ro+d\ro)=\frac12(d\ell)^2,
\eeq
where $d\ell$ is defined by the Fisher metrics \eqref{dlF} in the classical limit and
in the general quantum case it corresponds to the Fisher--KMB \eqref{dlKMB}
and not to the Fisher--Bures \eqref{dlB} metrics.
Consider a smooth (not necessarily geodesic) path $\gamma$  from $\ro$ to $\sio$.
We use $N$ reservoirs to transport $\ro$ into $\sio$
in $N$ steps. Consider a monotone sequence of intermediate states 
$\ro_i$ along the path, such that $\ro=\ro_0$ and $\ro_N=\sio$. Perform the transports 
$\ro\Rightarrow\ro_1\Rightarrow\ro_2\Rightarrow\dots\Rightarrow\ro_{N-1}\Rightarrow\sio$.
The total entropy production is the sum of the $N$ yields:
\beq
\Delta S=\sum_{i=0}^{N-1}S(\ro_i\Vert\ro_{i+1}).
\eeq
In the limit $N\rightarrow\infty$ each step $\Delta\ell_i$ between $\ro_i$ and $\ro_{i+1}$
can go to zero. In this asymptotic regime, applying eq. \eqref{relSdell} we can thus write   
\beq\label{DSl}
\Delta S=\frac12\sum_{i=0}^{N-1}(\Delta\ell_i)^2.
\eeq
This is an important expression but the r.h.s. is not yet the function of the
total path length $\ell_\gamma(\ro,\sio)$ while the lengths elements are constrained by it:
\beq
\sum_{i=0}^{N-1}\Delta\ell_i=\ell_\gamma(\ro,\sio).
\eeq
Keeping this constraint, let us minimize the entropy production \eqref{DSl}.
It is trivial that  each term on the r.h.s. must have the same value, i.e.,
$\Delta\ell_i=\ell_\gamma(\ro,\sio)/N$ for all $i$. This means
that the dense sequence $\ro_i$ must be evenly distributed along the path,
and the entropy production will be evenly distributed over the steps. The minimum
of the total entropy production along the given path $\gamma$ takes this form:
\beq\label{minDS}
\Delta S_\gamma = \frac{\ell_\gamma^2(\ro,\sio)}{2N}.
\eeq
The fastest approach to reversible transport is achieved on geodesics. 
In the classical case, inserting the geodesic length \eqref{Fl}  in  \eqref{minDS}, 
the ultimate lower bound on entropy production reads
\beq
\Delta S = \frac{2}{N}(\arccos F)^2.
\eeq
This expresses the desired sharp operational meaning of the classical 
Bhattacharyya fidelity \eqref{FC}.
The bounds are attainable in the limit $N\rightarrow\infty$ by operations on the system plus 
the reservoirs.

We find the simple meaning of the  Fisher--KMB length    
if we introduce the number $\nu$ of equilibrations per unit length, 
i.e., the density $\nu=N/\ell_\gamma$. Then
the entropy production becomes the following linear function of the path length:
\beq\label{DSlnu}
\Delta S_\gamma(\ell)=\frac{\ell_\gamma}{2\nu}.
\eeq
This suggests the following operational meaning of the  Fisher--KMB length.      
\emph{Consider the near-reversible state transport
via sequential equilibrations along a smooth path.
If $\nu\gg1$ is the density of equilibrations 
per unit path length then the entropy production 
is $(1/2\nu)$ per unit length.}

\section{Remarks and summary}
Paths in Fisher
geometries are perhaps the simplest non-local objects and, as such,
they can be interpreted as state transport processes. It has therefore seemed logical to 
look here for the  meaning of the geometries.  
One should mention the standard 
Kantorovich--Wasserstein theory \cite{kantorovich1960mathematical,vaserstein1969markov}
where the transport from distribution $p$ to $q$ happens via direct relocation of populations from
the $p_\al$'s into the $q_\beta$'s. The transport of quantum state $\ro$ into $\sio$,
too, can work similarly after a single unitary rotation of $\ro$ that makes it commute with $\sio$.    
This may seem much simpler than sequential equilibration. But the simple method
assumes the detailed control of the states in question. If it is not possible because,
e.g., the states are complicated many-body states then  the method of equilibration 
becomes more valuable.

Thermodynamic systems are the typical examples.
The Weinhold--Ruppeiner geometry \cite{weinhold1975metric,ruppeiner1979thermodynamics}
defines the thermodynamic distances between thermodynamic states and the thermodynamic
path lengths in general. 
It has long been known that the thermodynamic lengths coincide with statistical lengths
\cite{diosi1984metricization} and they are rooted in the the Fisher geometry of Gibbs states
\cite{janyszek1989geometrical}.
In thermodynamics, the expression $(\Delta\ell)^2/V$ quantifies 
how the statistical fluctuations of the thermodynamic
parameters tend to zero in the thermodynamic limit of infinite volume $V\rightarrow\infty$.
This does not interpret the global thermodynamic lengths $\ell$ but the local metrics. 
Studies of thermodynamic state transport
\cite{salamon1998geometry,salamon2023more,diosi2002shannon,diosi1996thermodynamic}
targeted  a relationship between the thermodynamic path length and the thermodynamic 
entropy production. 
Salamon and Nulton \cite{salamon1998geometry} recognized that in discrete sequential 
equilibration the equilibration rates cancel from the expression of entropy production
and they found the thermodynamic predecessor of eq. \eqref{minDS}. 
Scandi and Perernau-Llobet derived similar result in quantum thermodynamics, using 
Fisher--KMB geometry of quantum Gibbs states  \cite{scandi2019thermodynamic}.
Ref. \cite{diosi2002shannon} was the first attempt at constructing an
underlying classical microscopic mechanism of a specific irreversible state transport.
The present work constructed the abstract microscopic generalization valid in the 
Fisher--KMB geometries of arbitrary quantum states including the Fisher geometry special case 
of probability distributions. For this construction, the analytically tractable microscopic 
reservoir model and the related theorem, both proposed in ref. \cite{diosi2006exact}, 
have been instrumental. 

The global Fisher statistical distance both on space of probability distributions and on the
space of density  matrices with the KMB metrics got an operational interpretation for the 
first time. We obtain the sharp, not just qualitative, operational (physical) meaning
 of the classical Bhattacharyya fidelity. (Interestingly, the global Fisher--Bures geometry
 and the quantum fidelity are still waiting for their operational interpretations.)
The concrete choice of the  reservoir and the equilibration protocol is not likely to be
critical and alternative microscopic models should lead to the same 
relationships between entropy production and the Fisher--KMB geometry.
Future investigations may replace our stepwise  transport with some continuous protocol.

\begin{acknowledgments}
The author thanks Florio M. Ciaglia who  drew his attention to KMB metrics.     
This research was funded by 
the National Research, Development and Innovation Office (Hungary)
Frontline Research Excellence Program (Grant No. KKP133827).
\end{acknowledgments}

\bibliography{diosi2024}{}
\end{document}